# Steady 1D Stationary Currents of Spherical Gas Layer

Mikhail I. Ivanov[*]

**ABSTRACT**

Spherical layer of ideal gas is considered. The layer is in the sphere's gravity field. Existence possibility of steady 1D stationary currents of this layer is studied. This problem simulates zonal winds taking place in the atmospheres of some planets such as Venus, Titan, Jupiter and Saturn.

## INTRODUCTION

It is well known that the atmospheres of Venus, Titan, Jupiter, Saturn and some other planets have special dynamics in which zonal winds play main role. Zonal wind is a system of currents propagating along circles of latitude [1-3]. In the other hand, the main elements of large-scale atmospheric circulation of other planets (such as Mars) are the Hadley cells in which gas particles propagate neither latitudinal nor longitudinal directions [4]. Thus, atmospheres can be in qualitatively different dynamical regimes but the reason of this is not entirely clear.

Zonal winds are very steady; for example, Jovian zonal wind velocity profiles to be measured with a 17-year interval coincide with nearly graphic accuracy [2]. Zonal wind can be realized occasionally as an atmospheric superrotation when the planet's atmosphere rotates speeder than the planet itself. In the other hand, it is well known that only the solid-state rotations are steady in a closed system [5]. Obviously, the zonal wind existence and steadiness is maintained by a continuous energy inflow but a concrete mechanism of a system excitation is not ascertained yet.

It seems to be truly that the zonal wind is a stationary process that not depends on temporary factors such as tides and gravitational waves. Most obvious continuously acting factor is a latitude-disproportionate heating of the planet's surface (or the lower atmospheric layer in the case of gas planets). There exist several works based on this supposition (see, for example, [6, 7]). In [8] within the framework of this hypothesis the author has obtained results to be in a good agreement with astrometric data. However, the system of basic equations in [8] has been found underdeterminate and the author has supplemented it with the equation $\rho = \rho(r)$. But one can obtain a missing equation demanding a steadiness of the solutions studied. It has been done in the present work.

## FORMULATION OF THE PROBLEM AND SEARCH FOR 1D SOLUTIONS

We consider a free atmosphere dynamics quasi-stationary approximation. The corresponding system of the equations for the non-rotating coordinate system includes the equation of motion, the discontinuity equation, the equation for entropy flux and the equation of ideal gas [9]:

$$\frac{d\mathbf{v}}{dt} + \frac{1}{\rho}\nabla p + \nabla G = 0 \tag{1}$$

$$\frac{d\rho}{dt} + \rho \nabla \mathbf{v} = 0 \tag{2}$$

$$\rho \frac{dp}{dt} = \kappa p \frac{d\rho}{dt} + \frac{\varepsilon}{c_v} R\rho^2 T \tag{3}$$

$$p = R\rho T \tag{4}$$

where $G$ is the gravitational potential, $\kappa = c_p / c_v$ is the adiabatic number, equal to ratio of the specific heats, $R$ is the gas constant per unity of average molecular mass of gas, $\varepsilon$ is the non-adiabatic heat flux to unity of average molecular mass of gas per 1 K. Following [8], we will consider the atmosphere that have a rigid bottom – the planet's surface.

---

[*] Russian Academy of Sciences, A. Ishlinsky Institute for Problems in Mechanics, Moscow, Russia;
*e-mail*: m-i-ivanov@mail.ru

If we supposed $\varepsilon = 0$ and $G = G(r)$ where $r$ is the radial coordinate then one can demonstrate that the system (1)-(4) have 1D stationary solutions in the form of the currents propagating along circles of latitude. However, as shown in [8], the system (1)-(4) becomes underdeterminate by those assumptions. In [8] the system has been supplemented by the equation $\rho = \rho(r)$. The solutions obtained by this way have been found in a good agreement with astrometric data [8].

We suppose the gravitational potential to depend only on the radial coordinate $\partial G / \partial r = g(r)$. We consider stationary longitude-independent motions and express the temperature in (3) through the pressure and the density from (4). Then we have the basic system in the spherical coordinate system:

$$\frac{u}{r}\frac{\partial u}{\partial \theta} + w\frac{\partial u}{\partial r} + \frac{uw}{r} - \frac{v^2}{r}\cot\theta + \frac{1}{\rho r}\frac{\partial p}{\partial \theta} = 0 \tag{5}$$

$$\frac{u}{r}\frac{\partial v}{\partial \theta} + w\frac{\partial v}{\partial r} + \frac{uv}{r}\cot\theta + \frac{vw}{r} = 0 \tag{6}$$

$$\frac{u}{r}\frac{\partial w}{\partial \theta} + w\frac{\partial w}{\partial r} - \frac{u^2}{r} - \frac{v^2}{r} + \frac{1}{\rho}\frac{\partial p}{\partial r} + g(r) = 0 \tag{7}$$

$$\frac{1}{r\sin\theta}\frac{\partial}{\partial \theta}(\rho u \sin\theta) + \frac{1}{r^2}\frac{\partial}{\partial r}(\rho r^2 w) = 0 \tag{8}$$

$$\rho\left(\frac{u}{r}\frac{\partial p}{\partial \theta} + w\frac{\partial p}{\partial r}\right) = \kappa p\left(\frac{u}{r}\frac{\partial \rho}{\partial \theta} + w\frac{\partial \rho}{\partial r}\right) + \frac{\varepsilon}{c_v}p\rho \tag{9}$$

We will search for a solution of (4)-(9) in the form $\{\delta u, v_0 + \delta v, \delta w, \rho_0 + \delta\rho, p_0 + \delta p\}$, where $\delta u$, $\delta v$, $\delta w$, $\delta\rho$ and $\delta p$ are the small disturbances of the corresponding functions of the same order of smallness. Suppose the coefficient $\iota = \varepsilon / c_v$ characterizing the entropy flux to be the same order of smallness as disturbances $\delta u$, $\delta v$, $\delta w$, $\delta\rho$ and $\delta p$. Then the null approximation is:

$$-\frac{\rho_0 v_0^2}{r}\cot\theta + \frac{1}{r}\frac{\partial p_0}{\partial \theta} = 0 \tag{10}$$

$$-\frac{\rho_0 v_0^2}{r} + \frac{\partial p_0}{\partial r} + \rho_0 g(r) = 0 \tag{11}$$

As it was noticed in [8], the null approximation system (10)-(11) is underdeterminate. In [8] it was redefined by use of explicit setting of an additional equation for $\rho_0$. However, this additional equation can be obtained from the next (first) approximation. This approximation has a form:

$$-v_0^2 \delta\rho \cot\theta - 2\rho_0 v_0 \delta v \cot\theta + \frac{\partial \delta p}{\partial \theta} = 0 \tag{12}$$

$$\left(\frac{1}{r}\frac{\partial v_0}{\partial \theta} + \frac{v_0}{r}\cot\theta\right)\delta u + \left(\frac{\partial v_0}{\partial r} + \frac{v_0}{r}\right)\delta w = 0 \tag{13}$$

$$-\frac{2\rho_0 v_0}{r}\delta v + \frac{\partial \delta p}{\partial r} + \left(g(r) - \frac{v_0^2}{r}\right)\delta\rho = 0 \tag{14}$$

$$\frac{\rho_0}{r}\frac{\partial \delta u}{\partial \theta} + \rho_0 \frac{\partial \delta w}{\partial r} + \left(\frac{\rho_0}{r}\operatorname{ctg}\theta + \frac{1}{r}\frac{\partial \rho_0}{\partial \theta}\right)\delta u + \left(\frac{2}{r}\rho_0 + \frac{\partial \rho_0}{\partial r}\right)\delta w = 0 \tag{15}$$

$$\frac{1}{r}\left(\rho_0 \frac{\partial p_0}{\partial \theta} - \kappa p_0 \frac{\partial \rho_0}{\partial \theta}\right)\delta u = \left(\kappa p_0 \frac{\partial \rho_0}{\partial r} - \rho_0 \frac{\partial p_0}{\partial r}\right)\delta w + \upsilon p_0 \rho_0 \tag{16}$$

Note that the equations (13), (15) and (16) contain only two disturbances – $\delta u$ and $\delta w$. Thus, the system (13), (15) and (16) is overdeterminate in relation to $\delta u$ and $\delta w$. Joining it with the underdeterminate system (10)-(11), we obtain the completely determinate system of 5 equations with 5 unknown quantities – $v_0$, $\rho_0$, $p_0$, $\delta u$ and $\delta w$.

## INTEGRATION OF THE SYSTEM AND DISCUSSION

The system obtained includes the equations (10), (11), (13), (15) and (16). Its general integration is of great difficulty. In the other hand, it is obvious that an atmosphere is a thin layer where the radial coordinate $r$ has small changes. That is why we will search for a solution of this system near the planet's surface $r = a$. Enter the new variable $\xi = (r-a)/a$. For definiteness, we will neglect the contribution of atmospheric gas in the gravitational potential. Then, we have $g(r) = g_0 a^2 / r^2$. We will search for solutions in the form:

$$p_0(r,\theta) = \sum_{i=0} P_i(\theta)\xi^i \tag{17}$$

$$\rho_0(r,\theta) = \sum_{i=0} R_i(\theta)\xi^i \tag{18}$$

$$v_0(r,\theta) = \sum_{i=0} V_i(\theta)\xi^i \tag{19}$$

$$\delta u(r,\theta) = \sum_{i=0} U_i(\theta)\xi^i \tag{20}$$

$$\delta w(r,\theta) = \sum_{i=1} W_i(\theta)\xi^i \tag{21}$$

Here, $P_0$, $R_0$, $V_0$, $U_0$ are some distributions of the pressure, density and two velocity components at the surface $r = a$, the third (radial) velocity component equals to zero at the surface according to the impermeability condition. Substituting series (17)-(21) into the system (10), (11), (13), (15), (16) and equaling free terms to zero we obtain the next system of ODE's:

$$P_0' = V_0^2 R_0 \cot\theta \tag{22}$$

$$P_1 = R_0\left(V_0^2 - g_0 a\right) \tag{23}$$

$$V_0' + V_0 \cot\theta = 0 \tag{24}$$

$$R_0 W_1 + U_0 R_0' + R_0 U_0' + U_0 R_0 \cot\theta = 0 \tag{25}$$

$$U_0 R_0 P_0' - \kappa U_0 P_0 R_0' = \iota a P_0 R_0 \tag{26}$$

From (24) it follows $V_0 = C/\sin\theta$. Since all the functions entered must be limited we find $C = 0$ and, therefore, $V_0 = 0$. From here we obtain $P_0 = P_0^* = \text{const}$. Then, equaling terms each other at the first degree of $\xi$ we find:

$$P_1' = 0 \tag{27}$$

Designate $P_1 = P_1^* = \text{const}$. From (23) we obtain:

$$R_0 = -\frac{P_1^*}{g_0 a} \tag{28}$$

Then the equation (26) becomes:

$$\iota P_0^* P_1^* = 0 \tag{29}$$

whence follows $P_0 \equiv 0$ or $R_0 \equiv 0$. It is equivalent to the fact that the pressure or density at the planet's surface equals to zero that has no physical sense. Hence, we formulate the next statement: *the free atmosphere dynamics quasi-stationary approximation does not permit any steady 1D stationary currents to exist.*

In the other hand, the astrometric measurements show that the 1D currents most probably exist [1-3] and are steady [2]. Therefore, the statement obtained is a result of a non-adequate model choice. But how do we construct an adequate model for the zonal wind? Perhaps, we need to argue in this way: if $W_0 \neq 0$ then the statement obtained is not correct. For the giant planets like Jupiter and Saturn that has no solid surface this condition can be easily fulfilled but it is not so easy for the Earth-type planets like Venus and Titan. If there had been one more gas layer under studied one we might have renounced the condition $W_0 = 0$. It seems that it transfers difficulty to the lower layer but we now can use another model for it, for example, a viscous gas model. In fact, the planetary boundary layer of the Earth's atmosphere is under action of the unevenness of the Earth's surface [10] that is why it has to been considered as viscous. Therefore, the model constructed has to include a thin viscous planetary boundary layer just over the surface and an ideal gas layer over it where the 1D currents studied propagate. It is in accordance with meteorological data that show no 1D currents in the planetary boundary layer.

Return to the gas planets. Perhaps, these planets have a vertical convective flux in deep atmospheric layers and gas motions become horizontal only in the highest atmospheric regions, i.e. its zonal wind is an effect of its internal convection. If that is right then it gives us an opportunity to penetrate into giant planet upper atmospheric layers and study its convection and thermodynamics on the basis of the zonal wind measurements.

The author thanks V.G. Baydulov for fruitful discussions.
This work was supported by RFBR (project 11-01-00247).